\input{aipcheck}

\documentclass[
    ,final            
  ]
  {aipproc}

\layoutstyle{6x9}
\usepackage{rotating}

\begin{document}

\title{Coordinated AMBER and MIDI observations
 of the Mira variable RR Aql}

\classification{97.30.Jm, 97.10.Ex, 97.10.Me, 97.10.Sj}
\keywords      {techniques:interferometric - stars:AGB - stars:atmospheres - stars:mass-loss - stars:individual:RR Aql}

\author{Iva Karovicova}{
  address={European Southern Observatory, Germany, (ikarovic@eso.org, mwittkow@eso.org)}
}

\author{Markus Wittkowski}{
  address={European Southern Observatory, Germany, (ikarovic@eso.org, mwittkow@eso.org)}
}

\author{David A. Boboltz}{
  address={US Naval Observatory (dboboltz@usno.navy.mil)}
}

\author{\\*Michael Scholz}{
  address={Institut f$\ddot{u}$r Theoretische Astrophysik der Univ. Heidelberg}
  ,altaddress={ University of Sydney, Australia} 
}

\begin{abstract}
We have used near- and mid-infrared interferometry to investigate the pulsating atmosphere and the circumstellar environment of the Mira variable RR Aql. Observations were taken with the VLTI/AMBER (near infrared) and the VLTI/MIDI (mid infrared) instruments. We have obtained a total of 15 MIDI epochs between Apr 9, 2004 and Jul 28, 2007 covering 4 pulsation cycles and one AMBER epoch on Sep 9, 2006 at phase 2.82. This work is also part of an ongoing project of joint VLTI and VLBA observations to study the connection between stellar pulsation and the mass loss process. Here we present a comparison of the AMBER visibility data to a simple uniform disk model as well as to predictions by recent self-excited dynamic model atmospheres. The best fitting photospheric angular diameter of the model atmosphere at phase 2.82 is $\Theta$$_\mathrm{Phot}$= 9.9 $\pm$ 2.4 mas.
\end{abstract}

\maketitle


\section{Introduction}

Mira variables are cool luminous low-mass stars with large pulsation amplitudes and long periods on the AGB (Asymptotic Giant Branch). They exhibit strong mass loss with rates of up to 10$^{-7}$-$10^{-4}$ M$_\odot$/year, which is important for the evolution of the interstellar medium (ISM) and its chemical enrichment. These stars subsequently evolve toward the planetary nebula phase with a white dwarf as central star. 

Coordinated mutiwavelength observation (AMBER/MIDI/VLBA) is an efficient tool to better understand the mass loss process and its connection to the pulsation mechanism. While near-infrared observations provide information about the conditions near the stellar surface, mid-infrared observations are used to explore the characteristics of the molecular shells and the dust formation zone. VLBA observations allow us to add additional information about the properties of the environment using the maser radiation  that some of the most common molecules (SiO, H$_2$O, OH) emit. Previous results from this project of joint VLTI/VLBA observations for the Mira star S Ori can be found in Boboltz \& Wittkowski (2005) and Wittkowski et al. (2007).


\begin{table}
\caption{Observation on 09/09/2006}
\begin{tabular}{l l l l l l l}
\hline 
\hline
Target            &Purpose   &Wavelenght                   &Time      &$B_p$ [m] U1-U2/      &$PA_p$           \\
                &       &range[$\mu$m]                  &[UTC]         &U2-U3/U3-U1            &deg            \\
\hline
70 Aql         &Calibrator(K5 II) &                 & 1:53-2:06                   &53.1/44.5/96.8         &27.4/41.9/34.0    \\ 

RR Aql      &Science target  &2.12-2.20                        & 2:34-2:47               &55.6/46.5/101.4          &33.4/45.5/38.7 \\

RR Aql     &Science target   &2.12-2.20                          &  3:00-3:09              &56.1/46.6/102.1          &34.0/45.9/39.4  \\

70 Aql         &Calibrator(K5 II) &                   &   3:28-3:37                 &55.9/46.6/101.9          &33.6/45.7/39.1   \\

$\lambda$ Aqr      &Check star(M2 III) &2.19-2.27                         &3:59-4:08               &54.0/44.9/98.1          &25.4/39.9/32.0    \\

$\psi$ $_{01}$ Aql   &Calibrator(K1 III)&                & 4:27-4:36                     &54.5/45.3/99.0       &25.7/40.1/32.2     \\
\hline

\end{tabular}
\label{tab:LPer}
\end{table}

\begin{figure}
  \includegraphics[height=.5\textheight,angle=90]{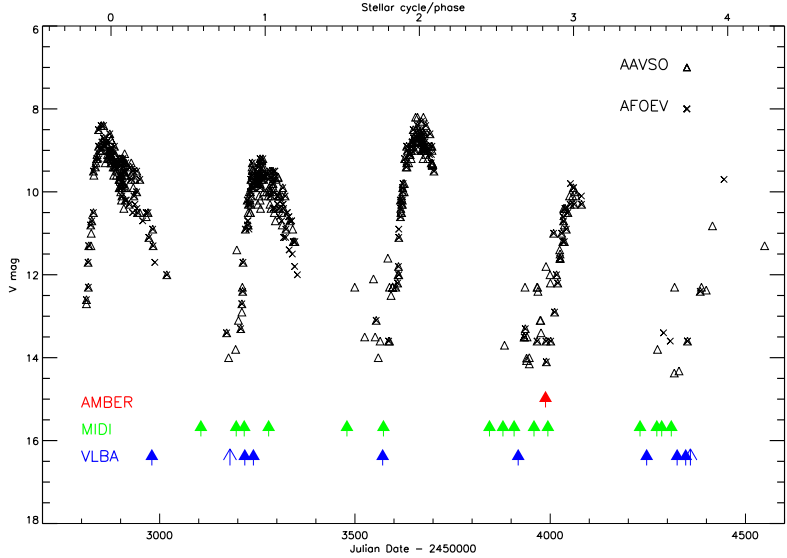}
 \caption{Visual light curve of RR Aql as a function of Julian Date and stellar cycle/phase. Data are from AAVSO and AFOEV databases. Period 394.78 days, the last maximum brightness T$_0$=2452875.4. The arrows indicate the dates of our VLTI/VLBA observations. AMBER - red arrow, MIDI - green arrows and VLBA - blue arrows (full arrows -SiO and simple - H$_2$O masers).}
\end{figure}

\section{Observation}

In this work we present interferometric observations which were obtained with the AMBER instrument in medium spectral resolution mode ($\Delta\lambda$/$\lambda$ $\sim$ 1500), using three Unit Telescopes (UTs) on 09/09/2006 (Julian day 245 3988). The observations are summarized in Table 1, which lists the name of the star and calibrators, the wavelength range, the time, the projected baseline lengths (B$_p$) and the position angle (PA$_p$) on the sky. The detector integration time (DIT) was 50 ms, the air-mass about 1.1, the optical seeing about 1.2 arcsec and the coherence time about 2.1 msec. The date of observation corresponds to a visual phase $\Phi$$_{vis}$= 2.82, with an uncertainty of about 0.1. The adopted limb-darkened diameters of 70 Aql and $\psi$ $_{01}$ Aqr are $\Theta$$_{LD}=$3.18$\pm$0.037 mas and $\Theta$$_{LD}=$2.18$\pm$0.025 mas respectively (Merand, Borde at al, 2005). We have used $\lambda$ Aqr as a check star to confirm the reliability of the instruments.

Figure 1 shows the visual lightcurve of RR Aql as a function of Julian day and stellar phase, based on values from the AAVSO (www.aavso.org) and AFOEV (http://cdsweb.u-strasbg.fr/afoev) databases. We used a period P = 394.78 (Samus et al. 2004) and JD of last maximum brightness T$_0$ = 245 2875.4 (Pojmanski 2002 - 2005). In the figure are also indicated the dates of our  VLTI and VLBA observations. For the data reduction we used the $\emph{Amdlib}$ package (version 2.1) with the yorick interface (provided by the Amber consortium and the Jean-Marie Mariotti  Center). 

\begin{figure}
  \includegraphics[height=.7\textheight,angle=90]{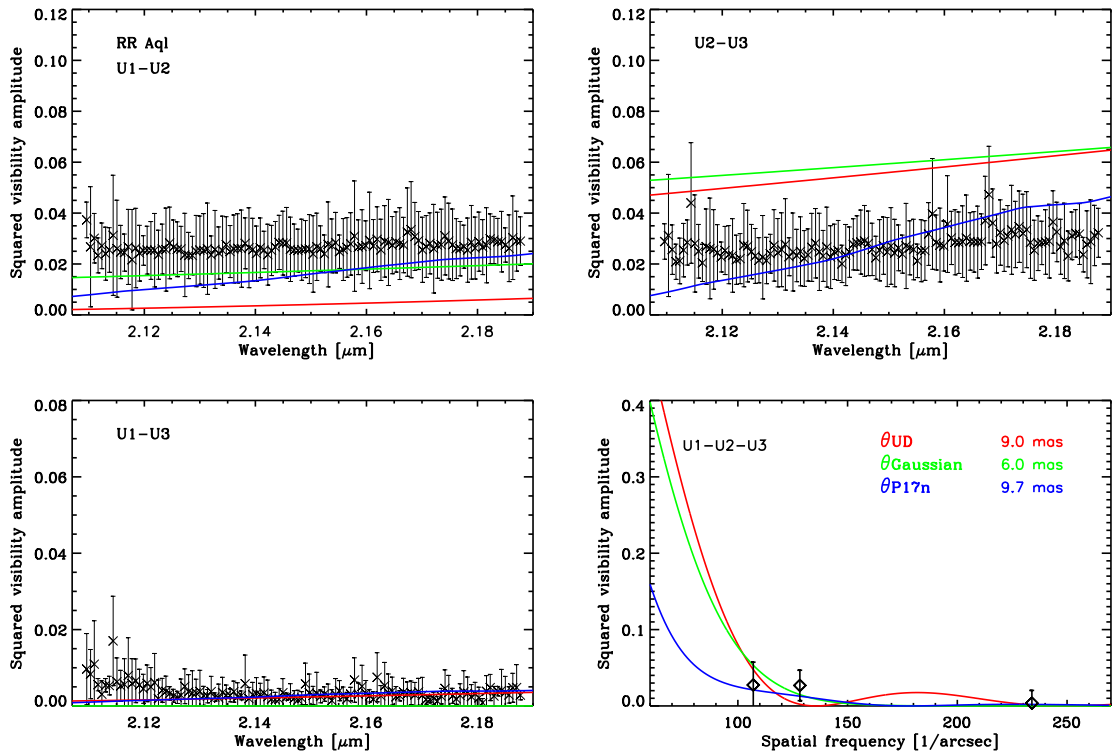}
 \caption{Measured visibility data of RR Aql compared to models of a UD (red line), of a Gaussian disk (green line), and of the P17n atmosphere model (blue line).}
\end{figure}

\begin{figure}
  \includegraphics[height=.7\textheight,angle=90]{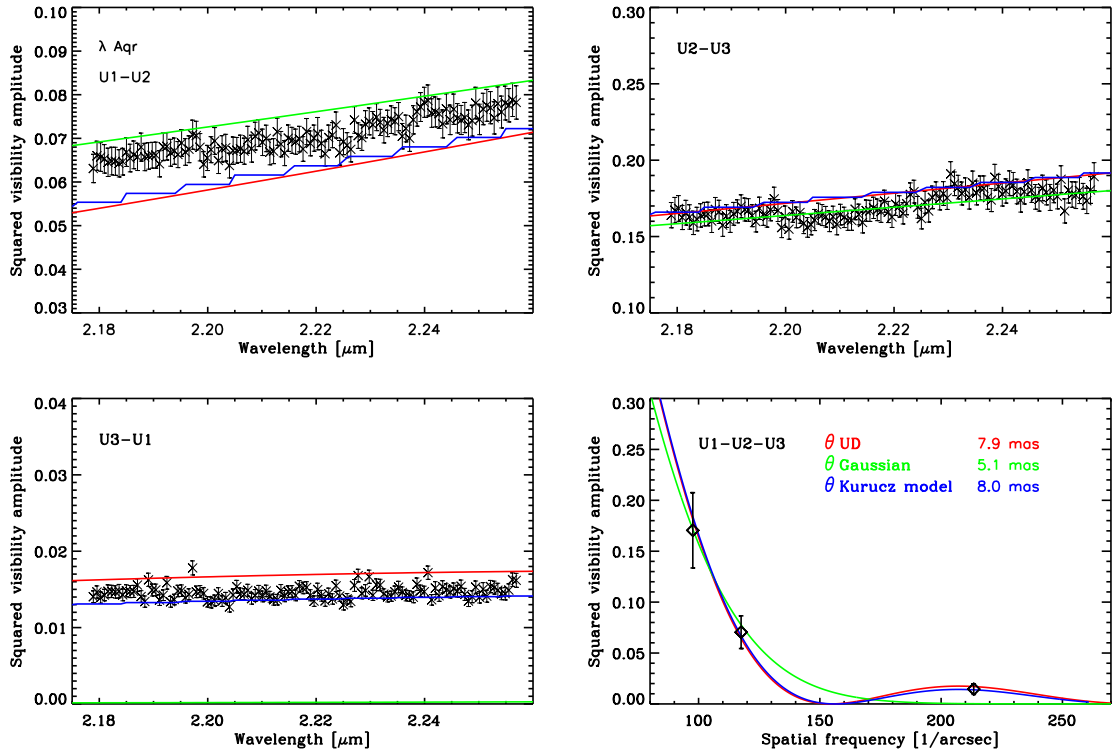}
 \caption{Measured visibility data of $\lambda$ Aqr compared to model of a UD (red line), of a Gaussian disk (green line), and of the ATLAS 9 atmosphere model (blue line).}
\end{figure}

\section{Results}
Figures 2 and 3 show our measured squared visibilities as a function of wavelength/spatial frequency. We compared the visibility data to the best fitting models of a uniform disk (UD), a Gaussian disk with a constant diameter and also to stellar atmospheres models. For RR Aql, we used the complete self-excited dynamic model atmospheres of Mira stars by Ireland et al.(2004 a,b). These models include the effect from molecular layers that lie above the continuum-forming photosphere. Out of the available phase and cycle combinations of the M and P series, the best fit to our measured AMBER visibility data its obtained with the model P17n. We estimate a continuum photospheric angular diameter of $\Theta$$_\mathrm{Phot}$= 9.9 $\pm$ 2.4 mas, based on an average of those models where | $\Phi$$_{vis}$ -  $\Phi$$_{model}$ | < 0.1.  
For $\lambda$ Aqr we used an ATLAS\,9 model atmosphere (Kurucz 1993). Note that the spectral resolution of the ATLAS 9 data is 0.01 m, compared to $\sim$ 0.0015 of our observations.


\section{Outlook}

We presented the recent results of AMBER observations of RR Aql, future work will be focused on comparing these AMBER results to mid-infrared interferometry (MIR) using our MIDI observations, and also to radio long-baseline interferometry (VLBA).



 




\bibliography{sample}

\IfFileExists{\jobname.bbl}{}
 {\typeout{}
  \typeout{******************************************}
  \typeout{** Please run "bibtex \jobname" to optain}
  \typeout{** the bibliography and then re-run LaTeX}
  \typeout{** twice to fix the references!}
  \typeout{******************************************}
  \typeout{}
 }

\end{document}